\documentclass[letterpaper]{article}%
\usepackage{aaai}
\usepackage{times}
\usepackage{helvet}
\usepackage{courier}
\usepackage{amssymb}
\usepackage{amsmath}
\usepackage{amsfonts}
\usepackage{graphicx}
\usepackage{epsfig}%
\providecommand{\U}[1]{\protect\rule{.1in}{.1in}}
\begin{document}

\title{Expressing Preferences using Preference Set Constraint Atoms}
\author{Alex Brik, Jeffrey B. Remmel\\Department of Mathematics, University of California, San Diego, CA 92093-0112\\\ }
\maketitle

\begin{abstract}
This paper introduces an extension of Answer Set Programming called Preference
Set Constraint Programming which is a convenient and general formalism to
reason with preferences. PSC programming extends Set Constraint Programming
introduced by Marek and Remmel \cite{MR} by introducing two types of
preference set constraint atoms, measure preference set constraint atoms and
pre-ordered preference set constraint atoms, which are extensions of set
constraint atoms. We show that the question of whether a PSC program has a
preferred stable model is CoNP-complete. We give examples of the uses of the
preference set constraint atoms and show that Answer Set Optimization
\cite{BNT03} and General Preference \cite{SP} can be expressed using
preference set constraint atoms.

\end{abstract}

\section{\bigskip Introduction}

The notion of a set constraint (SC)\ atom and a set constraint logic program
was introduced by Marek and Remmel in \cite{MR}. In this paper we extend these
notions to define preference set constraint (PSC) atoms and PSC logic
programs. The purpose of these extensions is to use PSC atoms to express preferences.

PSC\ programming is an intuitive and general formalism for expressing
preferences. We demonstrate its generality by showing that PSC programing can
be used to express optimal stable models of Answer Set Optimization (ASO) of
\cite{BNT03} and general preferences of \cite{SP}. An extension of PSC
programming can be used to express preferred answer sets and weakly preferred
answer sets of \cite{BrewkaE99}. However, due to space limitations, we will
not discuss the last two examples.

In this paper, we shall focus on the formal definitions of PSC programming.
However, there are a number of interesting issues concerning the best way to
implement PSC programs. While such issues are for the most part outside of the
scope of this paper, we note that an implementation of PSC\ programming will
not necessarily be a simple application of the definitions. The question of
what is the best way to implement PSC\ programming efficiently is one that
requires additional research, and we will only briefly discuss a few salient
implementation issues in this article.

The subject of expressing preferences using Answer Set Programming (ASP)\ has
been discussed extensively in the literature. Various approaches have been
proposed. We refer the reader to the article \cite{BNT08} for an accessible
overview of the subject and to a more detailed albeit older survey
\cite{DSTW04}.

In \cite{DSTW04} various approaches are classified for handling preferences in
nonmonotonic reasoning. The paper identified the following criteria.

\begin{itemize}
\item \textbf{Host system}. This is a particular formalism which is extended
to handle the preferences. In our case the host system is a SC logic program.

\item \textbf{What is the preference ordering an ordering on}?\ PSC
programming allows the user to directly specify preference orderings on sets
of atoms.

\item \textbf{Meta-level vs. object-level preferences}. This criteria
identifies whether preferences are imposed ``externally"\ on the host system,
or the preferences are used within the object theory. Our approach is a
meta-level approach.

\item \textbf{Static vs. dynamic preferences}. This criteria specifies whether
the preferences are fixed at the time that the theory is specified or can be
determined \textquotedblleft on the fly". PSC programming implements static preferences.

\item \textbf{Properties of the preference ordering}. PSC programming enforces
a pre-order on the set of PSC\ stable models.

\item \textbf{Prescriptive vs. descriptive preferences}. This criteria
concerns preference orderings on the rules and is not applicable to PSC programming.

\item \textbf{From preference to preferred results}. This criteria identifies
broad categories for methods that generate preferred answer sets from a theory
and a set of preferences. In PSC programming preferred PSC stable models are
standard PSC stable models satisfying additional criteria.
\end{itemize}

As noted in \cite{DSTW04}, the majority of ASP type systems that reason about
preferences use a preference ordering on the set of rules to express
preferences. An example of such an approach is given in \cite{BrewkaE99}.
There are exceptions such as for instance \cite{SI00}, \cite{Brewka02} that
specify preferences on the literals and more recently ASO which uses a second
preference program to specify preferences on the answer sets. ASO is in fact
the approach that is closest to PSC programming.

Delgrande et al. note that to have the most general ASP type system that can
reason about preferences, one must be able to handle preferences on sets of
objects. PSC programming allows the user to specify preferences on subsets of
atoms. This fact distinguishes PSC\ programming from most other approaches for
expressing preferences in ASP. PSC programming is different from ASO\ in that
the host system of PSC\ programming is a SC logic program, whereas the host
system of ASO\ is an ASP\ logic program. We will show that ASO programming can
be viewed as a particular case of PSC programming.

Since literals can be viewed as sets of cardinality 1, there are similarities
between PSC\ programming and some of the proposals that specify preferences on
the literals. Due to the space constraints we will limit the related
discussion to few remarks.

In \cite{Brewka02} Brewka introduces Logic Programs with Ordered Disjunction
(LPODs). The key feature of LPODs are ordered disjunctions of the form
$C_{1}\times...\times C_{n}$ in the heads of the rules. The meaning is that if
possible $C_{1}$, if not possible $C_{1}$ then $C_{2}$, ..., if not possible
$C_{1}$, ...,$C_{n-1}$ then $C_{n}$. The priorities among the answer sets of
LPODs are generated using the indices of the disjuncts included in the stable
models. As shown in \cite{BrewkaNS04} deciding whether $S$ is a preferred
answer set of LPOD\ $P$ is coNP-complete, which is the complexity of the same
problem for PSC\ programs. Thus a polynomial time translation between the
formalisms exists.

In \cite{BuccafurriLR00}, Buccafurri et al. extend the language of Disjunctive
Datalog by weak constraints (DATALOG$^{\vee,\lnot,c}$). The weak constraints
are constructs of the form $\Leftarrow L_{1},...,L_{n}$ where $L_{1}$, ...,
$L_{n}$ are literals. The weak constraints effectively allow to specify sets
of atoms, i.e. a weak constraint specifies those sets of atoms that violate
the constraint. $\Gamma$ is a model of a DATALOG$^{\vee,\lnot,c}$ program if
$\Gamma$ is a model of the underlying Disjunctive Datalog program and if it
minimizes the sum of the weights of the violated weak constraints.
DATALOG$^{\vee,\lnot,c}$ programs possess a certain similarity with the
measure PSC\ programs. In particular both specify preferences on sets and both
use numeric weights for the models. However, there are significant differences
between the two approaches. First, the host system of DATALOG$^{\vee,\lnot,c}$
is Disjunctive Datalog programs. For the measure PSC programs it is SC
programs. Second, integer weights associated with candidate models in
DATALOG$^{\vee,\lnot,c}$ allow the use of a binary search procedure in finding
models, whereas the use of binary search is not helpful for the measure PSC
programs since the stopping criteria is not easily determined due to the
arbitrary weights. Thus if the complexities of the host systems were the same,
the complexities of the search for optimal models would be different in the
two approaches.

The rest of the paper is structured as follows. In section 2, we will describe
the preliminaries of ASP\ and SC logic programming. In section 3, we will
introduce PSC programming. In the same section there will be a discussion
pertaining to implementing PSC programming. In section 4 we will show that
PSC\ programming can be used to express optimal stable models of ASO. In
section 5 we will show that the general preferences of \cite{SP} can be
expressed using PSC\ atoms. In section 6, we will discuss computational
complexity of PSC\ programming. In section 7, we will give conclusions and
directions for further research.

\section{Preliminaries}

\noindent\textbf{Answer Set Programming.}

Answer set programming is logic programming with stable model or answer set
semantics \cite{GL}, \cite{GL91}. ASP\ systems are ideal logic-based systems
to reason about a variety of types of data and integrate quantitative and
qualitative reasoning. The question of whether a finite propositional logic
program has a stable model is NP-complete \cite{Elkan}, \cite{MT}. It is also
the case that any NP search problem can be (uniformly) reduced to the problem
of finding a stable model of a finite propositional logic program \cite{MR2}.

A \emph{normal propositional logic program} $P$ consists of rules of the form
\[
C=a\leftarrow a_{1},...,a_{m},\;not\;b_{1},...,\;not\;b_{n}%
\]
where $a,a_{1},\ldots,a_{m},b_{1},\ldots,b_{n}$ are atoms and $not$ is a
non-classical negation operator. The set $prem\left(  C\right)  =\{a_{1}%
,$...$,a_{m}\}$ is called the set of \textit{premises} of rule $C$ and the set
$cons\left(  C\right)  =\{b_{1},\ldots,b_{n}\}$ is called the set of
\textit{constraints} of rule $C$. The atom $a$ is called the
\textit{conclusion} of rule $C$ and is denoted by $c\left(  C\right)  $.
Either $prem(C)$, $cons(C)$, or both may be empty. Let $H(P)$ denote the
Herbrand base of $P$. A subset $M\subseteq H(P)$ is called a model of a rule
$C$ if $prem(C)\subseteq M$ and $cons(C)\cap M=\emptyset$ implies $c(C)\in M$.
$M$ is a model of a program $P $ if it is a model of every rule of $P$.

Given $M\subseteq H(P)$, the Gelfond-Lifschitz transform $P^{M}$ of $P$ with
respect to $M$ is obtained by removing every rule $C$ such that $cons\left(
C\right)  \cap M\neq\emptyset$ and then removing the constraints from all the
remaining rules. $M$ is called a \textbf{stable model }of $P$ if $M$ is the
least model of $P^{M}$.\newline\ \newline\textbf{Set Constraint Logic
Programming.}

Set constraint programming was introduced in \cite{MR} as an extension of
$DATALOG^{\lnot}$. It generalized Answer Set Programming (ASP) with
cardinality constraint atoms or weight constraint atoms as defined in
\cite{NSS99}, \cite{NS00}.

Suppose that we are given a finite set of atoms $X$. We let $\mathcal{P}%
\left(  X\right)  $ denote the set of all subsets of $X$. A \emph{set
constraint atom} over $X$ is a pair $\left\langle X,F\right\rangle $ where
$F\subseteq\mathcal{P}\left(  X\right)  $. Given a set of atoms $M$ and a
SC\ atom $\left\langle X,F\right\rangle $, we say that $M$ satisfies
$\left\langle X,F\right\rangle $ (or $M$ is a model of $\left\langle
X,F\right\rangle $) and write $M\models\left\langle X,F\right\rangle $ if
$M\cap X\in F$.

A \emph{set constraint} (SC) rule $C$ is an expression of the form%
\[
s\leftarrow s_{1},...,s_{k}%
\]
where $s$, $s_{1}$, ..., $s_{k}$ are SC atoms. The set $body\left(  C\right)
=\left\{  s_{1},...,s_{k}\right\}  $ will be referred to as the body of the
rule $C$ and the atom $c\left(  C\right)  =s$ will be referred to as the
conclusion of the rule. A set of atoms $M$ is a \emph{model} of $C$ (or $M$
satisfies $C$) if $M\models s_{1}$, ..., $M\models s_{k}$ implies $M\models s
$. An SC program $P$ is a set of SC rules and a set of atoms $M$ is a model of
$P$ if $M$ is a model of every rule in $P$.

We note that the satisfaction of literals can be easily expressed in terms of
the satisfaction of set constraints. That is, for an atom $a$, $M\models a $
if and only $M\models\left\langle \left\{  a\right\}  ,\;\left\{  \left\{
a\right\}  \right\}  \right\rangle $ and $M\models not\ a$ if and only if
$M\models\left\langle \left\{  a\right\}  ,\;\left\{  \emptyset\right\}
\right\rangle $. Thus each literal $a$ or $not\;a$ in a normal logic program
can be written as a SC atom and hence each normal logic program can be
considered as SC logic program. However such a translation makes normal logic
programs harder to read. Thus, in what follows, we shall write $a$ for the SC
atom $\left\langle \left\{  a\right\}  ,\;\left\{  \left\{  a\right\}
\right\}  \right\rangle $ and $not\;a$ for a SC atom $\left\langle \left\{
a\right\}  ,\left\{  \emptyset\right\}  \right\rangle $.

Given an SC\ atom $\left\langle X,F\right\rangle $, the upper-closure
$\overline{F}$ of $F$ with respect to $X$ is the family $\overline
{F}=\{Y\subseteq X|\;\exists Z$ $\left(  Z\in F\text{ }\wedge\text{
}Z\subseteq Y\right)  \}$. A family $F$ of subsets of $X$ is \emph{closed} if
$\overline{F}=F$. Notice that the closure of a closed family $F$ of subsets of
$X$ is $F$ itself. The \emph{closure} of a SC atom $\left\langle
X,F\right\rangle $ is $\left\langle X,\overline{F}\right\rangle $.

A \emph{Horn SC\ rule} is a SC rule where the head of the rule is an ordinary
atom and all SC\ atoms in the body are closed, i.e. a rule of the form
\[
p\leftarrow\langle X,F_{1}\rangle,...,\langle X,F_{n}\rangle
\]
where for $i=1$, $2$, ..., $n$, $F_{i}=\overline{F_{i}}$. The reason for
calling such a rule Horn is that if $M$ satisfies a Horn SC\ rule $C$, then
all the supersets of $M$ will satisfy $C$. A \emph{Horn SC logic program}
(Horn SC\ program for short)\ is a SC program consisting entirely of Horn
SC\ rules. Given a Horn SC program $P$, we define the one-step provability
operator $T_{P}$ by letting $T_{P}(M)$ equal to the set of all $p$ such that
there exists a rule $p\leftarrow\left\langle X,F_{1}\right\rangle
$,...,$\left\langle X,F_{n}\right\rangle $ in $P$ where $M\models\langle
X,F_{i}\rangle$ for $i=1,\ldots,n$. It is easy to prove that the one-step
provability operator associated with a Horn SC program is monotone and hence a
Horn SC program $P$ has a least fixed point which is the smallest model of $P$.

The notion of SC stable model of a SC logic program is defined using a
modification of the Gelfond-Lifschitz transform called NSS transform. That is,
let $P$ be a SC program and let $M$ be a subset of atoms. The \emph{NSS
transform, }$NSS\left(  P,M\right)  $\emph{\ of }$P$\emph{\ with respect to
}$M $ is defined in two steps. First, eliminate from $P$ all rules whose
bodies are not satisfied by $M$. In the second step, for each remaining rule
$\left\langle X,F\right\rangle \leftarrow\left\langle X_{1},F_{1}\right\rangle
$, ..., $\left\langle X_{k},F_{k}\right\rangle $ and for each $a\in X\cap M$
generate the rule $a\leftarrow\left\langle X_{1},\overline{F_{1}}\right\rangle
$,..., $\left\langle X_{k},\overline{F_{k}}\right\rangle .$ The resulting
program $NSS\left(  P,M\right)  $ is a Horn SC program. Consequently, $NSS(P$,
$M)$ has a least model $N_{P,M}$. $M$ is called a \emph{SC stable model} of
$P$ if $M$ is a model of $P$ and $M=N_{P,M}$.

Marek and Remmel in \cite{MR} proved the following proposition showing the
equivalence of a normal logic program and its representation as SC program.

\textbf{Proposition 1.} Let $P$ be a normal logic program and let $M$ be a set
of atoms. Then $M$ is a stable model of $P$ in the sense of Gelfond and
Lifschitz if and only if $M$ is a stable model of $P$ viewed as SC program.

\section{Preference Set Constraints in Logic Programs}

In this section, we will define two types of preference set constraint atoms,
measure PSC atoms and pre-ordered PSC atoms, and two types of preference set
constraint programs, measure PSC programs and pre-ordered PSC programs which
are extensions of SC atoms and SC programs, respectively. We will also discuss
some of the issues related to the possible implementations of PSC programming.

A \emph{measure preference set constraint} (measure PSC) atom is a triple
$\left\langle X,F,\rho_{F}\right\rangle $ where $\langle X,F\rangle$ is a SC
atom and $\rho_{F}:F\rightarrow\lbrack-\infty,\infty]$ is a \emph{measure
function}. The SC reduct of $\left\langle X,F,\rho_{F}\right\rangle $,
$red(\langle X,F,\rho_{F}\rangle)$, is just the SC atom $\langle X,F\rangle$.
A \emph{pre-ordered preference set constraint} (pre-ordered PSC) atom is a
triple $\left\langle X,F,\leq_{F}\right\rangle $ where $\langle X,F\rangle$ is
a SC atom and $\leq_{F}$ is a pre-order on $F$. The SC reduct of $\left\langle
X,F,\leq_{F}\right\rangle $, $red(\langle X,F,\leq_{F}\rangle)$, is just the
SC atom $\langle X,F\rangle$. If $M$ is a set atoms, we say that $M$ is a
model of $\left\langle X,F,\rho_{F}\right\rangle $ if $M\models\langle
X,F\rangle$ and $M$ is a model of $\left\langle X,F,\leq_{F}\right\rangle $ if
$M\models\langle X,F\rangle$. If $\langle X,F\rangle$ is a SC atom, then we
let $red(\langle X,F\rangle)=\langle X,F\rangle$.

A measure PSC rule $C$ is a rule of the form
\[
s\leftarrow s_{1},\ldots,s_{k}%
\]
where $s_{1},\ldots,s_{k}$ are SC atoms and $s$ is either a SC atom or a
measure PSC atom. We define the $SC$ reduct of $C$, denoted by $red(C)$, to be
the rule $red(s)\leftarrow s_{1},\ldots,s_{k}$. A measure PSC program $P$ is a
set of measure PSC rules and we define the SC reduct of $P$ to be the set of
$red(C)$ such that $C$ is in $P$. We say that a set of atoms $M$ is a model of
$P$ if and only if $M$ is a model of $red(P)$ and $M$ is a stable model of $P$
if and only if it is a stable model of $red(P)$.

Similarly, a pre-ordered PSC rule $C$ is a rule of the form $s\leftarrow
s_{1},\ldots,s_{k}$ where $s_{1},\ldots,s_{k}$ are SC atoms and $s$ is either
a SC atom or a pre-ordered PSC atom. We define the $SC$ reduct of $C$, denoted
by $red(C)$, to be the rule $red(s)\leftarrow s_{1},\ldots,s_{k}$. A
pre-ordered PSC program $P$ is a set of pre-ordered PSC rules and we define
the SC reduct of $P$ to be the set of $red(C)$ such that $C$ is in $P$. Again
we say that a set of atoms $M$ is a model of $P$ if and only if $M$ is a model
of $red(P)$ and $M$ is a stable model of $P$ if and only if it is a stable
model of $red(P)$.

Our idea is that if we are given a measure PSC program or a pre-ordered PSC
program $P$, then the preference set constraint atoms can be used to induce a
pre-order on the set of stable models of $P$. Before we can talk about this
induced pre-order, we need to define a pre-order on the models of a set of PSC
atoms $T$. That is, suppose that $T$ is a set of pre-ordered PSC atoms and we
are given two sets of atoms $M_{1}$ and $M_{2}$ which satisfy every element of
$T$. Then we say that $M_{1}$\emph{\ is preferred to }$M_{2}$\emph{\ relative
to }$T$, written $T\models M_{1}\prec M_{2}$ if for all $\langle X,F,\leq
_{F}\rangle\in T$, $M_{1}\cap X\leq_{F}M_{2}\cap X$ and there is at least one
$\langle X,F,\leq_{F}\rangle\in T$ such that $M_{1}\cap X<_{F}M_{2}\cap X$.
(Here, as usual for two sets $A$ and $B$ and a pre-order $\leq$, $A<B$ denotes
$A\leq B$ and $B\not \leq A)$. We say that $M_{1}$\emph{\ is equivalent to
}$M_{2}$\emph{\ relative to }$T$, written $T\models M_{1}\sim M_{2}$ if for
all $\langle X,F,\leq_{F}\rangle\in T$, $M_{1}\cap X\leq_{F}M_{2}\cap X$ and
$M_{2}\cap X\leq_{F}M_{1}\cap X$. Hence, our pre-order on the models of $T$ is
essentially a product order over the set of local preference orders induced by
each of pre-ordered PSC atoms in $T$. We say that $M_{1}$\emph{\ is
indistinguishable from }$M_{2}$\emph{\ relative to }$T$, written $T\models
M_{1}\approx M_{2}$ if $T\not \models M_{1}\prec M_{2}$ and $T\not \models
M_{2}\prec M_{1}$.

A slightly weaker type of pre-order on models can be induced by measure PSC
atoms. That is, given a set $T$ of measure PSC atoms and two sets of atoms
$M_{1}$ and $M_{2}$ which are models of $T$, we say that $M_{1}$\emph{\ is
weakly preferred to }$M_{2}$\emph{\ relative to }$T$, written $T\models
M_{1}\prec_{w}M_{2}$ if
\begin{equation}
\sum_{\langle X,F,\rho_{F}\rangle\in T}\rho_{F}(M_{1}\cap X)<\sum_{\langle
X,F,\rho_{F}\rangle\in T}\rho_{F}(M_{2}\cap X). \label{wpref}%
\end{equation}
Note that for $M_{1}$ to be weakly preferred to $M_{2}$ relative to $T$, we do
not require that for every $\langle X,F,\rho_{F}\rangle\in T$, $\rho_{F}%
(M_{1}\cap X)\leq\rho_{F}(M_{2}\cap X)$, but only in the aggregate $M_{1}$ is
preferred to $M_{2}$. This type of pre-order induced on models of sets of
measure preference atoms allows the user more flexibility in specifying
preferences. This is because one is allowed to weigh local preferences so that
the weight coming from the $\rho_{F}$ associated with the measure PSC atom
$\langle X,F,\rho_{F}\rangle$ makes a much bigger contribution to
(\ref{wpref}) than the weight coming from the $\rho_{G}$ associated with the
measure PSC atom $\langle Y,G,\rho_{G}\rangle$. Thus it is possible to make
sure that the local preferences specified by $\langle X,F,\rho_{F}\rangle$ are
much more important than the local preferences specified by $\langle
Y,G,\rho_{G}\rangle$. We say that $M_{1}$\emph{\ is indistinguishable from
}$M_{2}$\emph{\ relative to }$T$, written $T\models M_{1}\approx_{w}M_{2}$ if
$T\not \models M_{1}\prec M_{2}$ and $T\not \models M_{2}\prec M_{1}$, i.e.
\[
\sum_{\langle X,F,\rho_{F}\rangle\in T}\rho_{F}(M_{1}\cap X)=\sum_{\langle
X,F,\rho_{F}\rangle\in T}\rho_{F}(M_{2}\cap X).
\]

We are now in position to define how we can use PSC programs to specify
preferences on stable models. We will start out considering what we call
\emph{simple PSC programs}. A simple pre-ordered PSC program is a pre-ordered
PSC program $P$ which consists of two types of rules:
\[
C_{1}=s\leftarrow s_{1},\ldots,s_{k}%
\]
where $s,s_{1},\ldots,s_{k}$ are SC atoms and
\begin{equation}
C_{2}=s\leftarrow\label{C2}%
\end{equation}
where $s$ is a pre-ordered PSC atom. Given a simple pre-ordered PSC program
$P$, we let $pref\left(  {P}\right)  $ denote the set of pre-ordered PSC atoms
that appear in a rule of type (\ref{C2}) in $P$. Note that any stable model
$M$ of $P$ must satisfy all the pre-ordered PSC atoms in $pref(P)$. Given two
stable models $M_{1}$ and $M_{2}$ of $P$, we say that $M_{1}$\emph{\ is
preferred to }$M_{2}$\emph{\ relative to }$P$, written $P\models M_{1}\prec
M_{2}$, if $pref(P)\models M_{1}\prec M_{2}$.

Similarly, a simple measure PSC program is a measure PSC program $P$ which
consists of two types of rules:
\[
D_{1}=s\leftarrow s_{1},\ldots,s_{k}%
\]
where $s,s_{1},\ldots,s_{k}$ are SC atoms and
\begin{equation}
D_{2}=s\leftarrow\label{D2}%
\end{equation}
where $s$ is a measure PSC atom. Given a simple measure PSC program $P$, we
let $pref\left(  {P}\right)  $ denote the set of measure PSC atoms that appear
in a rule of type (\ref{D2}) in $P$. Note that any stable model $M$ of $P$
must satisfy all the measure PSC atoms in $pref(P)$. Given two stable models
$M_{1}$ and $M_{2}$ of $P$, $M_{1}$\emph{\ is weakly preferred to }$M_{2}%
$\emph{\ relative to }$P$, written $P\models M_{1}\prec_{w}M_{2}$, if
$pref(P)\models M_{1}\prec_{w}M_{2}$.

\bigskip

To be practical, any implementation of PSC\ programming will have to be
restricted to a class of programs that can be represented in a compact way. In
this section we will discuss the \emph{algorithmic approach} for implementing
PSC\ programming - an approach that in many cases can produce compact
representations and that is conducive to efficient and practical
implementations of PSC\ programming.

The idea is that for a pre-ordered PSC atom $\left\langle X,F,\leq
_{F}\right\rangle $ or a measure PSC atom $\left\langle X,F,\rho
_{F}\right\rangle $ or a SC\ atom $\left\langle X,F\right\rangle $ $X$ and $F$
are implemented by algorithms $A_{X}$ and $A_{F}$ respectively. That is, for
any set of atoms $M$, the algorithm $A_{X}$ returns $M\cap X$, and, for any
set of atoms $M$, the algorithm $A_{F}$ returns $1$ if $M\in F$ and $0$ if
$M\notin F$. Thus an implementation of PSC rule may be a string of the form
$\left\langle X,F,q\right\rangle \leftarrow\left\langle X_{1},F_{1}%
\right\rangle ,...,\left\langle X_{n},F_{n}\right\rangle $, where $X$, $X_{1}%
$, ..., $X_{n}$ are names of the algorithms that for an input set of atoms $M$
return a subset of $M$. $F$, $F_{1}$, ..., $F_{n}$ are names of Boolean valued
algorithms that take a set of atoms as an input. For a pre-ordered PSC
program, $q$ will be a three valued algorithm that takes two sets of atoms as
inputs. For a measure PSC program, $q$ will be a real valued algorithm that
takes a set of atoms as an input. We note that an extension of ASP\ that uses
arbitrary algorithms was considered in \cite{HybridASP}.

In many cases this approach produces efficient implementations. Consider the
following example. Suppose that for a SC\ atom $\left\langle X,F\right\rangle
$, $F$ contains only those subsets of $X$ that consist of an even number of
elements. Implementation of this atom by enumerating $F$ is clearly
inefficient as there are $2^{\left\vert X\right\vert -1}$ sets with even
number of elements in $F$. Thus we do not want to represent $F$ by an
enumeration of all the subsets of $X$ of even cardinality. However, an
algorithm returning $1$ if and only if the input set has even number of
elements is trivial to implement and will run in $O\left(  |X|\right)  $ time.

An obvious implementation can be obtained as follows. Let $\{x_{1}%
,...,x_{n}\}$ be an enumeration of $X$. Thus any subset of $X$ can be
represented as a vector of $n$ binary digits (bits). For instance, suppose
that $X=\{x_{1}$, ..., $x_{8}\}$. Consider $Y\subseteq X$ where $Y=\{x_{1}$,
$x_{3}$, $x_{6}\}$. Then an 8 bit representation of $A$ is 101001. Assuming
this representation we have the following pseudocode for $A_{F}$.

{\small boolean AF (Y)\ }

{\small \ \ n = size(Y);}

{\small \ \ result = true;}

{\small \ \ for i=1:n if Y[i] == 1 then result = \symbol{126}result; endfor;}

\ {\small return result;}

{\small end}

where Y[i] is the ith bit of the bit vector Y, and \symbol{126}result is the
boolean NOT\ operation.

Suppose that instead of the subsets of $X$ of even cardinality we want to
enforce the cardinality constraint on the subsets of $Y$ of $X$ such that
$3\leq\left\vert Y\right\vert \leq\left\vert X\right\vert -3$. There are
$2^{\left\vert X\right\vert }-2(\left\vert X\right\vert \cdot\left(
\left\vert X\right\vert -1\right)  /2+\left\vert X\right\vert +1)$ such
subsets, and of course we do not want to enumerate them. However, as in the
previous example there is a simple algorithm that will implement this
cardinality constraint.

{\small boolean AF (Y)\ }

{\small \ \ n = size(Y);}

{\small \ \ cardinality = 0;}

{\small \ \ for i=1:n if Y[i] == 1 then cardinality = cardinality +\ 1;
endfor;}

{\small \ \ return 3
$<$%
= cardinality \&\&\ cardinality
$<$%
= n-3;}

{\small end}

PSC\ semantics specifies what an implementation of PSC\ programming should do.
It is not a prescription for how PSC\ programming should be implemented. Thus
the fact that PSC programming deals with sets should not be understood to mean
that PSC programming has to be implemented by enumerating sets. Efficient
implementations of PSC programming will not use such an approach. We will use
two examples to motivate the fact that semantics does not necessarily
prescribe how the formalism is to be implemented.

First, consider stable model semantics. The formalism shows that a stable
model can be found by choosing a subset of the Herbrand base of a logic
program and then checking that the subset is a stable model. Now, while
undoubtedly useful, the formalism is impractical if implemented as stated.
Indeed, modern ASP\ solvers such as \emph{smodels }\cite{SimonsNS02} and
\emph{clasp }\cite{GebserKNS07} use efficient algorithms to implement stable
model semantics, where the algorithms do not rely on searching through the
entire powerset of the Herbrand base of a logic program, which would be the
case if the semantics was considered as a prescription for implementations.

Second, consider cardinality constraint programming \cite{NSS99}. The
formalism has been implemented in \emph{smodels-2} \cite{Simons99} and is
generally considered to be a practical extension of ASP. However, its
practicality follows from the fact that efficient implementations of
cardinality constraint programming exist.

To illustrate how the algorithmic approach for implementing PSC programming
might work we will consider the problem of finding a vertex cover of size less
than $K$ for a given graph, with a preference for the covers that include
vertex $w$.

For a given graph a measure PSC program can be constructed as follows. For
every edge $\left(  u,v\right)  $ include a rule$\left\langle \left\{
u,v\right\}  ,\;\left\{  \left\{  u\right\}  ,\left\{  v\right\}  \right\}
\right\rangle \leftarrow$, specifying that either $u$ or $v$ is in a stable
model. Also include the following rule $\left\langle V,F,\rho_{F}\right\rangle
\leftarrow$, where $V$ is the set of all the vertices of the graph and for
$U\subseteq V$ $U\in F$ if $\left\vert U\right\vert <K$ and
\[
\rho_{F}=\left\{
\begin{array}
[c]{l}%
0\text{ if }w\in U\\
1\text{ if }w\notin U
\end{array}
\right.  .
\]

Now, an algorithmic implementation of $V$ is very simple: for any subset $U$
of the Herbrand base $V\left(  U\right)  =U$.

The implementation of $F$ is very similar to the cardinality constraint
implementation above. Finally the algorithm roF(U) simply has to check the
condition U[iw]==1, where iw is the index in the bit array corresponding to
the atom $w$. That is $x_{iw}=w$. The preference can be given by any of the
three approaches for measure PSC programs.

The example demonstrates that compact representations of PSC\ programs can be
created and that such representations are conducive to efficiently
implementing PSC\ programming.

\bigskip

In many cases, simple PSC programs are adequate to express preferences. In
fact pre-ordered simple PSC programs will be used to express optimal stable
models of ASO.

We will now proceed with the examples illustrating how PSC\ programming can be used.

\textbf{Example 1.}\label{Ex1} Bob is a Ph.D. student who is about to graduate
from his university. Bob is guessing that he will have multiple job offers and
wants to determine a method by which he will make his decision. Bob identifies
two important criteria in making a decision. These are the type of institution
and its location. He thus introduces the following atoms: R - for the job at a
research university, T - for the job at a teaching university, C - for the job
in a company, CAL-for the job located in California, and NCal-for the job not
located in California. Thus the set of atoms is $X=\{R$, $T$, $C$, $CAL$,
$NCAL\}$. Then any offer is described by a set of atoms from the following
family of sets: $F=\{\{A$, $B\}$ $|$ $A\in\{R$, $T$, $C\}$, $B\in\{CAL$,
$NCAL\}\}$. Finally Bob decides on the following ordering of sets $\leq_{F}$:
$\{R$, $CAL\}$\ $<_{F}$ $\{R$, $NCAL\}$ $<_{F}$ $\{T$, $CAL\}$ $<_{F}$\ $\{T$,
$NCAL\}$ $<_{F}$ $\{C$, $CAL\}$ $<_{F}\{C$, $NCAL\}$.

In this example and all the following examples in this section, we will assume
that Bob is trying to decide between two jobs $j_{1}$ and $j_{2}$. Thus we
introduce a base program $P_{0}$. $P_{0}$ contains the rule
\[
\langle\{j_{1},j_{2}\},\{\{j_{1}\},\{j_{2}\}\}\rangle\leftarrow.
\]
This rule says that, in any stable model $M$, exactly one of $j_{1}$ and
$j_{2}$ is contained. Then we add rules to specify the relevant information
about jobs $j_{1}$ and $j_{2}$. For example, if $j_{1}$ is a job at a research
university in California and $j_{2}$ is a job at a teaching university outside
of California, we would add the following rules: $R\leftarrow j_{1}%
$,$\ \ \ Cal\leftarrow j_{1}$, $T\leftarrow j_{2}$,$\ \ \ NCal\leftarrow
j_{2}$.

Then Bob's preferences can be described by the simple pre-ordered PSC\ program
$P_{1}$ which consists of $P_{0}$ plus the rule $\left\langle X,F,\leq
_{F}\right\rangle \leftarrow$. \ \ $\square$

\textbf{Example 2.}\label{Ex2} Bob soon realizes that not all locations
outside of California have the same weight. He is actually more likely to
consider an offer from a location which is near California than from a
location which is far from California. Bob thinks that a job at a research
university is preferable to a job at a teaching university and that a job at a
teaching university is preferable to a job at a company. Yet, Bob notices that
he will prefer a job from a teaching university in California to a job from a
research university which is more than 500 miles away from California.

He thus revises his original approach. There are still predicate atoms $R$,
$T$, $C$ to specify a research university, or a teaching university or a
company respectively. However, now Bob introduces a set of predicate atoms
$\widehat{D}=\{D\left(  x\right)  |x \in\mathbb{N}\}$, where $\mathbb{N}$ is
the set of natural numbers and $D\left(  x\right)  $ indicates a distance $x$
from California. Thus $D\left(  0\right)  $ indicates that the location of the
job is in California. Note that in an implementation of PSC programming
$\widehat{D}$ can be an algorithm that on an input $M$ will simply return the
set of atoms in $M$ of the form $D(x)$ for $x \in\mathbb{N}$. While
$\widehat{D}$ represents an infinite set, its implementation can be compact
and efficient.

Now let $Z=\{R$, $T$, $C\}\cup\widehat{D}$ and let $H=\{\{A$, $B\}|$ $A\in
\{R$, $T$, $C\}$, $B\in\widehat{D}\}$. Bob defines a measure function as
follows $\rho_{H}\left(  \left\{  A,D\left(  x\right)  \right\}  \right)
=\tau\left(  A\right)  +x$ where $\tau\left(  R\right)  =0$, $\tau\left(
T\right)  =500$ and $\tau\left(  C\right)  =1000$.

Thus Bob's preferences are specified by the simple measure PSC\ program which
consists of $P_{0}$ plus the rule $\left\langle Z,H,\rho_{H}\right\rangle
\leftarrow.$ \ $\square$

If we consider more general pre-ordered PSC programs $P$ and measure PSC
programs, then we have several natural choices for how to induce a pre-order
on the set of stable models of $P$. If $P$ is a pre-ordered (measure)\ PSC
program and $M$ is a stable model of $P$, then we let $pref\left(  P,M\right)
$ denote the set of all pre-ordered (measure)\ PSC\ atoms $s$ such that there
is a rule $C=s\leftarrow s_{1},\ldots,s_{k}$ where $s$ is a pre-ordered
(measure) PSC atom and $M$ satisfies the body of $C$. Since all stable models
of $P$ are models of $P$ by definition, $M$ must be a model of $pref(P,M)$.

Note, however, that if $M_{1}$ and $M_{2}$ are stable models of $P$ it is not
necessarily the case that $pref(P,M_{1})=pref(P,M_{2})$. Now in the case where
$pref(P,M_{1})=pref(P,M_{2})$, the obvious thing to do for the pre-ordered PSC
programs is to say that $M_{1}$ is preferred to $M_{2}$ relative to $P$ if and
only if $pref(P,M_{1})\models M_{1}\prec M_{2}$. However, if $pref(P,M_{1}%
)\neq pref(P,M_{2})$, then one has several natural choices. First, one can
simply consider the pre-ordered PSC atoms in $pref(P,M_{1})\cap pref(P,M_{2}%
)$, i.e. the pre-ordered PSC atoms which are the conclusions of rules of $P$
which are satisfied by both $M_{1}$ and $M_{2}$. Thus we say that $M_{1}$ is
\emph{in common preferred} to $M_{2}$ relative to $P$, written $P\models
M_{1}\prec_{ic}M_{2}$, if and only if $pref(P,M_{1})\cap pref(P,M_{2})\models
M_{1}\prec M_{2}$. A second natural choice that one might want to use in
certain situations is to take the point of view that satisfying a pre-ordered
PSC atom $s$ that appears in the head of a rule in $P$ is more preferable than
not satisfying $s$. Thus we say that $M_{1}$ is \emph{in total preferred} to
$M_{2}$ relative to $P$, written $P\models M_{1}\prec_{it}M_{2}$, if and only
if either (a) $pref(P,M_{1})\supset pref(P,M_{2})$ and either $pref(P,M_{1}%
)\cap pref(P,M_{2})\models M_{1}\prec M_{2}$ or $pref(P,M_{1})\cap
pref(P,M_{2})\models M_{1}\sim M_{2}$ or (b) $pref(P,M_{1})=pref(P,M_{2})$ and
$pref(P,M_{1})\cap pref(P,M_{2})\models M_{1}\prec M_{2}$.

\textbf{Example 3.}\label{Ex3} As Bob has more time to contemplate the job
offers, he realizes that his life will be simplified if the job is in a town
where there is a good public transportation system. Also, being a classical
music lover, Bob considers an easy access to live classical concerts as one of
the factors in making his decision. Not being sure about the weight that
public transportation and live classical music concerts should have in the
decision making process, he reverts to the pre-ordered PSC\ program from
Example 1. Bob reasons that he will keep his preferences as they already are,
except when there is an access to live classical music concerts in the area.
In that case, a location outside of California with a good system of public
transportation is preferable to a location in California without such a
system. Thus Bob introduces two new predicate atoms:\ $CM$ - to indicate the
presence of local access to live classical music concerts and $PT$ to indicate
the presence of a good system of public transportation. Bob adds a new rule to
the program $P_{1}$ to produce a new program $P_{3}$. $P_{3}$ consists of
$P_{0}$ plus the following two rules:
\[
\left\langle X,F,\leq_{F}\right\rangle \leftarrow\text{ \ \ \ }\left\langle
Y,G,\leq_{G}\right\rangle \leftarrow CM
\]
where $Y=\{CAL$, $NCAL$, $PT\}$, $G=\{\{CAL$, $PT\}$, $\{NCAL$, $PT\}$,
$\{CAL\}$, $\{NCAL\}\}$ and $\{CAL$, $PT\}<_{G}\{NCAL$, $PT\}$ $<_{G}$
$\{CAL\}$ $<_{G}$ $\{NCAL\}$.

Now $P_{3}\models M_{2}\prec_{ic}M_{1}$ since $pref\left(  P_{3},M_{1}\right)
$ $\cap$ $pref\left(  P_{3},M_{2}\right)  $ $=$ $\{\langle X$, $F$, $\leq
_{F}\rangle\}$ and $\langle X$, $F$, $\leq_{F}\rangle\models$ $M_{2}\prec
M_{1}$. However $P_{3}\models M_{1}\approx_{it}M_{2}$ because $pref\left(
P_{3},M_{1}\right)  $ $\supset pref\left(  P_{3},M_{2}\right)  $ and $\langle
X$, $F$, $\leq_{F}\rangle\models M_{2}\prec M_{1}.\square$\newline

Similarly we can define a pre-order on the set of stable models of measure
PSC\ programs. In the case where $pref(P,M_{1})=pref(P,M_{2})$, the obvious
thing to do is to say that $M_{1}$ is weakly preferred to $M_{2}$ relative to
$P$ if and only if $pref(P,M_{1})\models M_{1}\prec_{w}M_{2}$. However, if
$pref(P,M_{1})\neq pref(P,M_{2})$, then one has several natural choices. One
is to simply consider the measure PSC atoms in $pref(P,M_{1})\cap
pref(P,M_{2})$, i.e. the measure PSC atoms which are the conclusions of rules
of $P$ which are satisfied by both $M_{1}$ and $M_{2}$. Thus we say that
$M_{1}$ is \emph{in common weakly preferred} to $M_{2}$ relative to $P$,
written $P\models M_{1}\preceq_{w,ic}M_{2}$, if and only if $pref(P,M_{1})\cap
pref(P,M_{2})\models M_{1}\prec_{w}M_{2}$. As before, our second natural
choice is to take the point of view that satisfying a measure PSC atom $s$
that appears in the head of a rule in $P$ is more preferable than not
satisfying $s$. Thus we say that $M_{1}$ is \emph{in total weakly preferred}
to $M_{2}$ relative to $P$, written $P\models M_{1}\prec_{w,it}M_{2}$, if and
only if (a) $pref(P,M_{1})\supset pref(P,M_{2})$ and either $pref(P,M_{1})\cap
pref(P,M_{2})\models M_{1}\prec_{w}M_{2}$ or $pref(P,M_{1})\cap pref(P,M_{2}%
)\models M_{1}\approx_{w}M_{2}$ or (b) $pref(P,M_{1})=pref(P,M_{2})$ and
$pref(P,M_{1})\cap pref(P,M_{2})\models M_{1}\prec_{w}M_{2}$. We also have a
third natural choice that one might want to use in certain circumstances which
is just to compare the two sums $\sum_{\langle X,F,\rho_{F}\rangle\in
pref(P,M_{1})}\rho_{F}(M_{1}\cap X)$ and \newline$\sum_{\langle X,F,\rho
_{F}\rangle\in pref(P,M_{2})}\rho_{F}(M_{2}\cap X)$. Thus we say that $M_{1}$
is \emph{in sum weakly preferred} to $M_{2}$ relative to $P$, written
$P\models M_{1}\prec_{w,is}M_{2}$, if
\[
\sum_{\langle X,F,\rho_{F}\rangle\in pref(P,M_{1})}\rho_{F}(M_{1}\cap X)<
\]%
\[
\sum_{\langle X,F,\rho_{F}\rangle\in pref(P,M_{2})}\rho_{F}(M_{2}\cap X).
\]

\textbf{Definition 1. }\textit{A set of atoms }$M$\textit{\ is called an in
common preferred PSC stable model of a pre-ordered PSC program }%
$P$\textit{\ if }$M$\textit{\ is a PSC stable model of }$P$\textit{\ and for
all PSC stable models }$M^{\prime}$\textit{\ of }$P$\textit{, }$P\not \models
M^{\prime}\prec_{ic}M$\textit{.}

An \emph{in total preferred PSC stable model}, an \emph{in common weakly
preferred PSC stable model}, an \emph{in total weakly preferred PSC stable
model}, an \emph{in sum weakly preferred PSC stable model }are defined
similarly. To refer to any of these definitions without explicitly naming them
we may say that $M$ \emph{is a preferred PSC stable model of a PSC program}
$P$.

\section{PSC Programs and Answer Set Optimization Programs}

As was stated in the introduction, the closest approach to PSC\ programming is
ASO. In ASO, one starts with an ASP\ program $P_{gen}$ over a set of atoms
$At$ and then a preference specification is given by a separate preference
program $P_{pref}$. The rules of $P_{pref}$ are of the form%
\begin{equation}
C_{1}>...>C_{k}\leftarrow a_{1}\text{, ..., }a_{n}\text{, }not\text{ }%
b_{1},...,\;not\;b_{m} \label{preference rule}%
\end{equation}
where $a_{i}$s and $b_{j}$s are literals and the $C_{i}$s are Boolean
combinations over $At$. Here a Boolean combination over $At$ is a formula
built of atoms in $A$ by means of disjunction, conjunction, strong negation
$\lnot$ and default negation $not$, with the restriction that strong negation
is allowed to appear only in front of atoms, and default negation is allowed
to appear only in front of literals.

Next suppose that we are given a set of literals $S$. Then the definition of
$S$ satisfying a Boolean combination $C$, written $S\models C$, uses the
standard inductive definition of satisfaction of propositional formulas except
that $S\models not$ $l$ where $l$ is literal if and only if $l\notin S$. Then
we define the satisfaction degree $v_{S}\left(  r\right)  $ for any rule of
the form of (\ref{preference rule}) by setting (i) $v_{S}\left(  r\right)  =I$
if either the body of $r$ is not satisfied by $S$ or the body of $r$ is
satisfied by $S$, but none of the $C_{i}$s are satisfied and (ii)
$v_{S}\left(  r\right)  =\min\{i:$ $S\models C_{i}\}$ if the body of $r$ is
satisfied by $S$ and at least one of the $C_{i}$s is satisfied by $S$. This
allows one to define a satisfaction vector $V_{S}=(v_{S}\left(  r_{1}\right)
$, ..., $v_{S}\left(  r_{n}\right)  )$ for any answer set $S$ of $P_{gen}$ for
a preference program $P_{pref}=\{r_{1}$, ..., $r_{n}\}$.

One can then use satisfaction vectors to define a pre-order on answer sets of
$P_{gen}$ as follows. First for any two possible values $a$ and $b$ of
$v_{S}\left(  r\right)  $ (i) $a\geq b$ if $a=I$ and $b=1$ or if $a=1$ and
$b=I$, (ii) $a>b$ if $a=I$ and $b$ $\in$ $\{2$, $3$, ...$\}$, and (iii) $a>b$
if $a$, $b$ $\in$ $\{1$, $2$, ...$\}$ and $a<b$ relative to the usual order on
the natural numbers. Then for two sets of literals $S_{1}$ and $S_{2}$
$V_{S_{1}}\geq V_{S_{2}}$ if $v_{S_{1}}\left(  r_{i}\right)  \geq v_{S_{2}%
}\left(  r_{i}\right)  $ for every $i\in\{1$, ..., $n\}$ $V_{S_{1}}>V_{S_{2}}$
if $V_{S_{1}}\geq V_{S_{2}}$ and for some $i\in\{1$, ..., $n\}$ $v_{S_{1}%
}\left(  r_{i}\right)  $ $>$ $v_{S_{2}}\left(  r_{i}\right)  $. $S_{1}\geq
S_{2}$ if $V_{S_{1}}\geq V_{S_{2}}$ and $S_{1}>S_{2}$ if $V_{S_{1}}>V_{S_{2}}%
$. Finally, a set of literals $S$ is an \emph{optimal model of an
ASO\ program} $(P_{gen}$, $P_{pref})$ if $S$ is an answer set of $P_{gen}$ and
there is no answer set $S^{\prime}$ of $P_{gen}$ such that $S^{\prime}>S$.

We will now show how optimal models of an ASO program can be expressed using
PSC programming. Let $\left(  P_{gen}\text{, }P_{pref}\right)  $ be an
ASO\ program.

Let $At$ be the set of all atoms that occur in the rules of $P_{gen}$ and
$P_{pref}$. Note that any stable model of $A$ of $P_{gen}$ must be a subset of
$At$. Consequently, any optimal model of $(P_{gen}$, $P_{pref})$ must be a
subset of $At$.

Let $P_{pref}$ consist of the rules $W_{1}$, $W_{2}$, ..., $W_{\gamma}$ where
$W_{i}$ is of the form $C_{1}^{i}>...>C_{k}^{i}\leftarrow a_{1}^{i}$, ...,
$a_{n_{i}}^{i}$, $not$ $b_{1}^{i},...,\;not\;b_{m_{i}}^{i}$.

We will now construct a simple pre-ordered PSC\ program $P$ and we will show
that there is a one-to-one correspondence between the preferred PSC\ stable
models of $P $ and the optimal stable models of $(P_{gen}$, $P_{pref})$.

The Herbrand base $H\left(  P\right)  $ of $P$ will consist of $At$, the set
$\overline{At}=\{\overline{a}|$ $a\in A\}$ where for $a\in At$ $\overline{a}$
is a new atom not in $At$, and where for all $a$, $b\in At$ if $a\neq b$ then
$\overline{a}\neq\overline{b}$, and a new atom $D$ so that $D\notin
At\cup\overline{At}$. That is $H\left(  P\right)  =At\cup\overline{At}%
\cup\left\{  D\right\}  $.

For a set $A\subseteq At$ define $\overline{A}=\{\overline{a}|$ $a\in A\}$.
For a rule $W_{i}\in P_{pref}$ let $At\left(  W_{i}\right)  $ be the set of
all the atoms that occur in $W_{i}$.

We define a PSC program $P_{pref}^{\prime}$ to be the set of all rules
\[
\left\langle \overline{At\left(  W_{i}\right)  },\;\overline{\mathcal{P}%
\left(  At\left(  W_{i}\right)  \right)  },\;\leq_{i}\right\rangle
\leftarrow.
\]
The pre-order $\leq_{i}$ is defined as follows. For $A\subseteq At$,
$B\subseteq At$ $\overline{A}\leq_{i}\overline{B}$ if one of the following
conditions hold.

1. $A\not \models body\left(  W_{i}\right)  $; 2. $A\models body\left(
W_{i}\right)  $ and $A\not \models c\left(  W_{i}\right)  $; 3. $A\models
body\left(  W_{i}\right)  $ and $A\models C_{1}^{i}$; 4. $A\models body\left(
W_{i}\right)  $ and $A\models C_{z}^{i}$ where $z$ is minimal and $B\models
body\left(  W_{i}\right)  $ and $B\models C_{j}^{i}$ where $j$ is minimal and
$z\leq j$, i.e., $\overline{A}\leq_{i}\overline{B}$ iff $v_{A}\left(
W_{i}\right)  \geq v_{B}\left(  W_{i}\right)  $.

Let $P_{cons}$ be the set of rules consisting of two rules for each $a\in At$
\ $\overline{a}\leftarrow a$ \ and $D\leftarrow\overline{a}$, $not\;a$,
$not\;D$.

Then the PSC program $P$ is defined by $P=P_{gen}\cup P_{pref}^{\prime}\cup
P_{cons}$.

We can prove the following two theorems.

\textbf{Theorem 1. }\textit{1. For all }$B\subseteq At,$\textit{\ if }%
$B$\textit{\ is a stable model of }$P_{gen}$\textit{\ in the sense of Gelfond
and Lifschitz, then }$B\cup\overline{B}$\textit{\ is a stable model of }%
$P$\textit{\ in the set of PSC.}

\textit{2. For all }$A\subseteq H\left(  P\right)  $, \textit{\ if }%
$A$\textit{\ is a stable model of }$P$\textit{\ in the sense of PSC, then }$A=
B \cup\overline{B}$ where $B$ \textit{\ is a stable model of }$P_{gen}%
$\textit{\ in the sense of Gelfond and Lifschitz.}

\textbf{Theorem 2. }\textit{1. If }$B$\textit{\ is an optimal stable model of
}$\left(  P_{gen}\text{, }P_{pref}\right)  $, \textit{\ then }$B\cup
\overline{B}$\textit{\ is a preferred PSC\ stable model of }$P$\textit{.}

\textit{2. If }$A$\textit{\ is a preferred PSC\ stable model of }%
$P$,\textit{\ then }$A\cap At$\textit{\ is an optimal stable model of
}$\left(  P_{gen}\text{, }P_{pref}\right)  $\textit{.} \newline

A reasonable concern in the construction of $P$ is the efficiency of the
implementation of the PSC\ atoms $\left\langle \overline{At\left(
W_{i}\right)  },\;\overline{\mathcal{P}\left(  At\left(  W_{i}\right)
\right)  },\;\leq_{i}\right\rangle $.The implementation of such an atom can be
a string $\left\langle X,\text{ }F,\;O\right\rangle $ where $X$, $F$, $O$ are
the names of the algorithms. The algorithm corresponding to $X$ on the input
set of atoms $M$ will return $\overline{At\left(  W_{i}\right)  }\cap M$. The
algorithm corresponding to $F$ will always return $1$ since any subset of
$\overline{At\left(  W_{i}\right)  }\cap M$ is in $\overline{\mathcal{P}%
\left(  At\left(  W_{i}\right)  \right)  }$. The algorithm corresponding to
$O$ on the inputs $\overline{A}$ and $\overline{B}$ will return $1$ iff
$\overline{A}\leq_{i}\overline{B}$ and $0$ otherwise. The algorithm for $O$
can be as efficient as the algorithm in the implementation of ASO. Our
algorithm can use the ASO\ algorithm to check the conditions $A\models
body\left(  W_{i}\right)  $, $A\models c\left(  W_{i}\right)  $, $B\models
body\left(  W_{i}\right)  $, $A\models C_{z}^{i}$, $B\models C_{z}^{i}$ that
are necessary for its evaluation.

\section{Using PSC\ Programs to Express General Preferences}

In this section we will show that the general preferences of \cite{SP} can be
expressed using PSC\ atoms. In \cite{SP}, the language $\mathcal{PP}$ for
planning preferences specification was introduced. $\mathcal{PP}$ allows users
to elegantly express multi-dimensional preferences among plans that achieve
the same goal. In $\mathcal{PP}$, users can define \emph{general preferences}
by building them from simpler \emph{atomic preferences} using a small number
of special operators.

A \emph{basic desire formula} (basic desire for short) is a formula expressing
a single preference about a trajectory. Son and Pontelli provide a formal
definition of the notion of \emph{basic desire}, as well as a formal
definition of the notion of a \emph{trajectory}, and define what it means for
a trajectory \emph{to satisfy} a basic desire. For the purposes of this paper,
it is not necessary to restate these definitions. It will suffice to assume
that a basic desire is a formula $\phi$ in some language and that, for any
trajectory $\alpha$, we can determine whether $\alpha$ \emph{satisfies} $\phi
$, written $\alpha\models\phi$, or whether $\alpha$\emph{\ does not satisfy
}$\phi$, written $\alpha\not \models \phi$. We will also assume that each
basic desire corresponds to a unique predicate atom. In addition, we will use
the basic desires and the predicate atoms corresponding to them interchangeably.

Let $\phi$ be a basic desire formula and let $\alpha$ and $\beta$ be two
trajectories. The trajectory $\alpha$ is \emph{preferred} to the trajectory
$\beta$, written $\alpha\prec_{\phi}\beta$, if $\alpha\models\phi$ and
$\beta\not \models \phi$. $\alpha$ and $\beta$ are \emph{indistinguishable
with respect to }$\phi$, written $\alpha\approx_{\phi}\beta$, if either
(1)$\ \alpha\models\phi$ and $\beta\models\phi$ or (2)\ $\alpha\not \models
\phi$ and $\beta\not \models \phi$.

An \emph{atomic preference formula} is defined as a formula of the type
$\phi_{1}\lhd\phi_{2}\lhd$...$\lhd\phi_{n}$ where $\phi_{1}$, ..., $\phi_{n}$
are basic desire formulas. If $\alpha$ and $\beta$ are trajectories and
$\psi=\phi_{1}\lhd\phi_{2}\lhd$...$\lhd\phi_{n}$ is an atomic preference
formula, then we say $\alpha$ and $\beta$ are \emph{indistinguishable with
respect to }$\psi$ (written as $\alpha\approx_{\psi}\beta$) if $\forall i$
$(1\leq i\leq n$ $\Rightarrow$ $\alpha\approx_{\phi_{i}}\beta)$ and $\alpha$
is \emph{preferred to }$\beta$ with respect to $\psi$ (written as $\alpha
\prec_{\psi}\beta$) if $\exists\left(  1\leq i\leq n\right)  $ such that
$\forall\left(  1\leq j<i\right)  $ $\alpha\approx_{\phi_{j}}\beta$ and
$\alpha\prec_{\phi_{i}}\beta$.

The set of \emph{general preference formulas} (general preferences) are
defined via the following inductive definition: (i) every atomic preference
formula $\psi$ is a general preference formula, (ii) if $\psi_{1}$, $\psi_{2}
$ are general preference formulas, then $\psi_{1}\&\psi_{2}$, $\psi_{1}%
|\psi_{2}$, and $!\psi_{1}$ are general preference formulas, and (iii) if
$\psi_{1}$, $\psi_{2}$, ..., $\psi_{k}$ are general preference formulas, then
$\psi_{1}\lhd\psi_{2}\lhd$ . . . $\lhd\psi_{k}$ is a general preference formula.

Let $\psi$ be a general preference formula and let $\alpha$, $\beta$ be two
trajectories. Then we say $\alpha$ is \emph{preferred to }$\beta$\emph{\ with
respect to }$\psi$ (written $\alpha\prec_{\psi}\beta$) if:

1. $\psi$ is an atomic preference formula and $\alpha\prec_{\psi}\beta$

2. $\psi=\psi_{1}\&\psi_{2}$ and $\alpha\prec_{\psi_{1}}\beta$ and
$\alpha\prec_{\psi_{2}}\beta$

3. $\psi=\psi_{1}|\psi_{2}$ and ($\alpha\prec_{\psi_{1}}\beta$ \ and
$\alpha\approx_{\psi_{2}}\beta$) or ($\alpha\approx_{\psi_{1}}\beta$ and
$\alpha\prec_{\psi_{2}}\beta$) or ($\alpha\prec_{\psi_{1}}\beta$ and
$\alpha\prec_{\psi_{2}}\beta$).

4. $\psi=!\psi_{1}$ and $\beta\prec_{\psi_{1}}\alpha$

5. $\psi=\psi_{1}\lhd\psi_{2}\lhd$ . . . $\lhd\psi_{k}$, and there exists
$1\leq i\leq k$ such that ($\forall\left(  1\leq j<i\right)  $ $\alpha
\approx_{\psi_{j}}\beta$ and $\alpha\prec_{\psi_{i}}\beta$) \newline

We say $\alpha$ is \emph{indistinguishable from }$\beta$\emph{\ with respect
to }$\psi$ (write $\alpha\approx_{\psi}\beta$) if:

1. $\psi$ is an atomic preference formula and $\alpha\approx_{\psi}\beta$

2. $\psi=\psi_{1}\&\psi_{2}$, $\alpha\approx_{\psi_{1}}\beta$ and
$\alpha\approx_{\psi_{2}}\beta$

3. $\psi=\psi_{1}|\psi_{2}$, $\alpha\approx_{\psi_{1}}\beta$ and
$\alpha\approx_{\psi_{2}}\beta$

4. $\psi=!\psi_{1}$ and $\alpha\approx_{\psi_{1}}\beta$

5. $\psi=\psi_{1}\lhd\psi_{2}\lhd$ . . . $\lhd\psi_{k}$, and for all $1\leq
i\leq k$ $\alpha\approx_{\psi_{i}}\beta$. \newline

Let $\Delta$ be the set of all the basic desire predicate atoms. Let $X$ be a
finite subset of $\Delta$ and $F\subseteq P\left(  X\right)  $. Let $\leq_{F}$
be a pre-order on $F$. For a trajectory $\alpha$ let $\Delta\left(
\alpha\right)  \equiv\left\{  d\in\Delta|\;\alpha\models d\right\}  $ i.e. the
set of all the basic desires satisfied by $\alpha$. Then we say that $\alpha
$\emph{\ satisfies }$\left\langle X,F,\leq_{F}\right\rangle $ if
$\Delta\left(  \alpha\right)  \cap X\in F$. We say that a trajectory $\alpha
$\emph{\ is preferred to a trajectory }$\beta$\emph{\ with respect to
}$\left\langle X,F,\leq_{F}\right\rangle $ if $\left\langle X,F,\leq
_{F}\right\rangle \models\Delta\left(  \alpha\right)  \prec\Delta\left(
\beta\right)  $. This will be denoted by $\left\langle X,F,\leq_{F}%
\right\rangle \models\alpha\prec\beta$. We say that trajectories $\alpha
$\emph{, }$\beta$\emph{\ are indistinguishable with respect to }$\left\langle
X,F,\leq_{F}\right\rangle $ if $\left\langle X,F,\leq_{F}\right\rangle
\models\Delta\left(  \alpha\right)  \approx\Delta\left(  \beta\right)  $. This
will be denoted by $\left\langle X,F,\leq_{F}\right\rangle \models
\alpha\approx\beta$.

\textbf{Theorem 3. }\textit{For a general preference formula }$\phi$
\textit{there exists a pre-ordered PSC\ atom }$\left\langle X_{\phi},F_{\phi
},\leq_{\phi}\right\rangle $\textit{ such that for two trajectories }$\alpha
$\textit{, }$\beta$\textit{, }$\alpha\prec_{\phi}\beta\;$\textit{\ in the
sense of \cite{SP} iff }$\left\langle X_{\phi},F_{\phi},\leq_{\phi
}\right\rangle \models\alpha\prec\beta$\textit{ and }$\alpha\approx_{\phi
}\beta\;$\textit{in the sense of \cite{SP} iff }$\left\langle X_{\phi}%
,F_{\phi},\leq_{\phi}\right\rangle \models\alpha\approx\beta$\textit{.}

The proof, omitted due to space limitations shows how to construct
$\left\langle X_{\phi},F_{\phi},\leq_{\phi}\right\rangle $.

\section{\label{complexity}The Computational Complexity of the Set of
Preferred PSC Stable Models}

We can prove the following result on the computational complexity of finding a
\textquotedblleft preferred\textquotedblright\ PSC stable models of a PSC
program. Please note that the theorem's preconditions are formulated to be
useful for analyzing algorithmic implementations of PSC programming as
discussed in section 3.

\textbf{Theorem 4. }\textit{Let }$P$\textit{\ be a finite PSC program. Let
}$At$\textit{\ be the set of all atoms appearing in }$P$\textit{\ and assume
that }$At$\textit{\ is finite. Suppose that for any subset }$M$\textit{\ of
}$At$\textit{\ and any PSC atom }$\left\langle X,F,\leq_{F}\right\rangle
$\textit{\ or }$\left\langle X,F,\rho_{F}\right\rangle $\textit{\ or any SC
atom }$\left\langle X,F\right\rangle $\textit{\ determining the intersection
}$M\cap X$\textit{\ and the membership }$M\cap X\in F$\textit{\ can be done in
polynomial time on the number of elements in }$M$\textit{\ and }$X$\textit{.
If }$P$\textit{\ is a pre-ordered PSC program, then suppose that for any PSC
atom }$\left\langle X,F,\leq_{F}\right\rangle $\textit{\ in }$P$\textit{\ for
any two sets in }$F$\textit{, the comparison operation }$\leq_{F}$\textit{\ on
these subsets can be performed in polynomial time on the number of elements in
the subsets. If }$P$\textit{\ is a measure PSC program, then suppose that for
any PSC atom }$\left\langle X,F,\rho_{F}\right\rangle $\textit{\ in }%
$P$\textit{\ and any subset }$A$\textit{\ of }$F$\textit{, the computation
}$\rho_{F}\left(  A\right)  $\textit{\ can be performed in polynomial time on
the number of elements of }$A$\textit{. Then given }$M\subseteq At$\textit{,
the problem of determining whether }$M$\textit{\ is a preferred PSC stable
model of }$P$\textit{\ for any of the induced pre-orders on stable models
described in Section 3 is CoNP-complete relative to the size of }%
$P$\textit{\ and the number of elements in }$At$\textit{.} \newline

Now, in \cite{BNT03} it was shown that the problem of deciding whether there
exists an answer set for an ASO program $P=(P_{gen}$, $P_{pref})$ is
NP-complete and the problem of deciding whether $S$ is an optimal model of $P
$ is coNP-complete. This is the complexity of the corresponding problems for
PSC programs. That is, the problem of deciding whether there exists a stable
model for a PSC program $P$ is NP-complete and the problem of deciding whether
a PSC stable model $M$ of a PSC program $P$ is a preferred PSC stable model is
co-NP complete.

\section{Conclusions and directions for further research}

In this paper, we have introduced an approach to specifying preferences in ASP
called Preference Set Constraint Programming which is an extension of SC
programming of \cite{MR}. PSC programming uses two types of PSC
atoms:\ pre-ordered PSC atoms and measure PSC atoms. These atoms can be used
to define pre-ordered PSC\ programs and measure PSC\ programs. For pre-ordered
PSC programs, we have considered two approaches for specifying preferences:
\textquotedblleft in common"\ and \textquotedblleft in total". For the measure
PSC programs we have considered three approaches:\ \textquotedblleft in
common", \textquotedblleft in total" and \textquotedblleft in sum". We show
that the problem of determining whether $M$ is a preferred PSC stable models
of a PSC program is CoNP-complete. To demonstrate the expressive power of PSC
programming, we have shown that PSC\ programming can be used to express
optimal stable models of ASO \cite{BNT03}, and the general preferences of Son
and Pontelli \cite{SP}. It is also the case that the preferred stable models
and the weakly preferred stable models of \cite{BrewkaE99} can be expressed by
an extension of PSC programming, although it was not discussed in this paper.

We have only briefly discussed implementations of PSC\ programming, but
clearly it is an important issue for applications.

There are a number of areas for further research on PSC programming. One is to
study the exact relationships between PSC programming and other approaches for
reasoning with preferences. In particular, one must pay special attention to
the efficiency of expressing preferences in any two systems that are being
compared. A second is to study extensions of PSC programming where we add the
ability of the user to express preference by preference orderings on rules. We
will study both of these questions in subsequent papers.

Finally, the five approaches for specifying preferences using PSC programs can
be viewed as special cases of the following generalization. A
\emph{PSC\ system} $R$ is a pair $\left\langle P,\leq\right\rangle $ where $P$
is a PSC program and $\leq$ is a pre-order on the set of PSC\ stable models of
$P$ with the property that if $M_{1}$ and $M_{2}$ are PSC stable models of $P$
such that $pref\left(  P,M_{1}\right)  $ $=$ $pref\left(  P,M_{2}\right)  $
and $M_{1}\leq M_{2}$, then $pref\left(  P,M_{1}\right)  \models M_{1}\prec
M_{2}$ or $pref\left(  P,M_{1}\right)  \models M_{1}\sim M_{2}$ if $P$ is a
pre-ordered PSC$\ $program and $pref\left(  P,M_{1}\right)  \models M_{1}%
\prec_{w}M_{2}$ or $pref\left(  P,M_{1}\right)  \models M_{1}\approx_{w}M_{2}$
if $P$ is a measure PSC\ program. We suggest that one should study abstract
properties of PSC systems.

\bibliographystyle{aaai}
\bibliography{bibliography}

\end{document}